\documentclass[oribibl, final]{llncs}

\usepackage{cite}
\usepackage[pdftex]{graphicx}

\usepackage{subfig}
\usepackage{url}
\usepackage{hyperref}

\newcommand{\citep}{\cite}

\begin{document}

\title{Modeling and Execution of Multienterprise Business Processes}

\author{Robert Singer\inst{1} \and Johannes Kotremba\inst{2} \and Stefan Ra\ss\inst{2}}

\institute{FH JOANNEUM -- University of Applied Sciences \and StrICT Solutions GmbH}

%\email{\{robert.singer\}@fh-joanneum.at}

\maketitle

\begin{abstract}
We discuss a fully featured multienterprise business process plattform (ME-BPP) based on the concepts of agent-based business processes. Using the concepts of the subject-oriented business process (S-BPM) methodology we developed an architecture to realize a platform for the execution of distributed business processes. The platform is implemented based on cloud technology using commercial services. For our discussion we used the well known Service Interaction Patterns, as they are empirically developed from typical business-to-business interactions. We can demonstrate that all patterns can be easily modeled and executed based on our architecture.
\end{abstract}

\section{Introduction}
\label{introduction}

Actual developments in business and technology driven new business models foster more than ever the need for mature business process management (BPM) methodologies and corresponding supporting technologies for \emph{distributed} business processes, so called choreographies. Business processes cannot be seen as isolated workflows for administrative purposes, but as a means to \emph{coordinate a value system} with supply chain partners. Communication is the very nature of a business process choreography -- or, in other words, any choreography is a set of \emph{structured communication patterns}. That means a choreography defines how work is done, taking into account all involved organizations. Distributed execution of a business process means that every process participant may use its own process execution engine. The overall process is then executed by interconnecting multiple engines. The engines may even also run on a mobile device.

This demand is reflected in new developments in the domain of BPM, such as \emph{BPM Platform as a Service} (bpmPaaS), \emph{multi-enterprise Business Process Platform} (ME-BPP), \emph{Cloud BPM}, and \emph{Social BPM}. The term bpmPaaS can be defined~\citep{Anonymous:2012tf} as ``\emph{the delivery of BPM platform capabilities as a cloud service by a service provider}''. A ME-BPP is defined~\citep{Anonymous:2012tf} as ``\emph{high-level conceptual model of a multistakeholder environment, where multi-enterprise applications are operated. multi-enterprise applications are those that are purposely built to support the unique requirements for business processes that span more than one business entity or organization. They replace multiple business applications integrated in serial fashion}''.

For a distributed execution of processes, two important prerequisites are needed: a \emph{suitable process modeling} technique and a \emph{flexible communication platform}~\citep{Aitenbichler:2011lr}. As elaborated in the following section, we have chosen the agent based approach to model a distributed system. To be more specific, we build on the Subject-oriented BPM methodology, as defined in~\citep{Fleischmann:2012va}.

To implement a communication platform, we need an architecture, which includes a graphical business process and\slash or rule modeling capability, a process registry\slash repository to handle the modeling metadata and a process execution and either a state management engine and\slash or a rule engine as minimal request. To realize a bpmPaaS and\slash or ME-BPP system a cloud infrastructure is needed to model and execute processes which \emph{span across more than one business entity or organization}.

\section{Subject-oriented BPM}
\label{subject-orientedbpm}

\begin{figure}[htbp]
\centering
\includegraphics[keepaspectratio,width=3.5in,height=0.75\textheight]{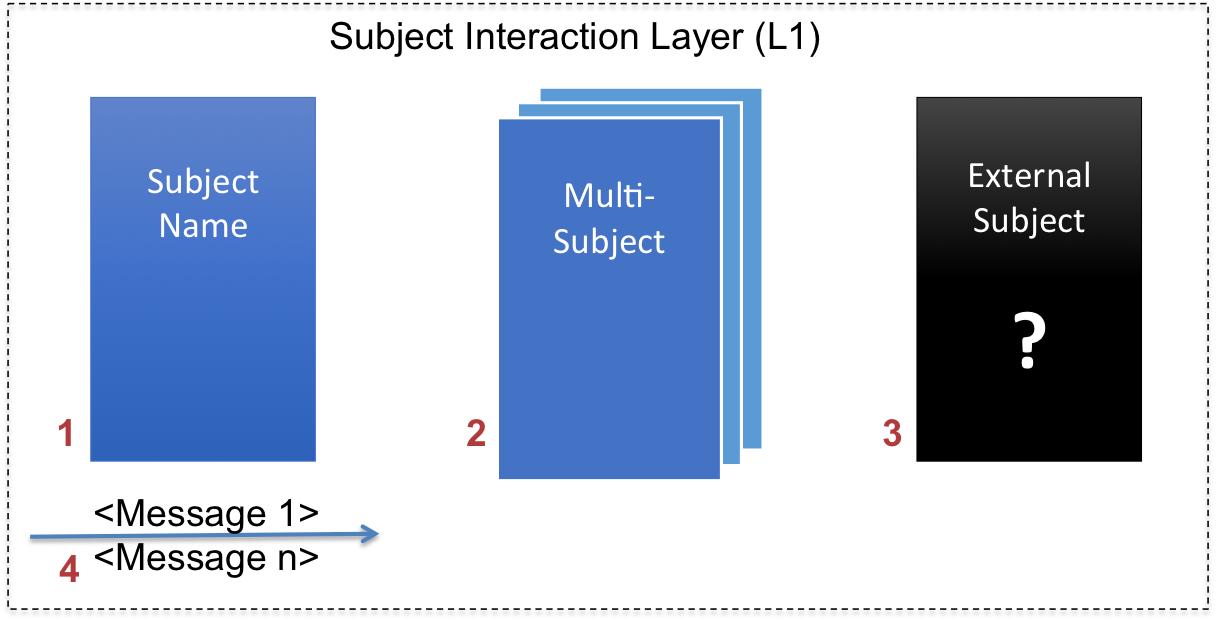}
\caption{Supported S-BPM Language Elements of the \emph{Subject Interaction Diagram} (Layer 1).}
\label{layer1}
\end{figure}

\begin{figure}[htbp]
\centering
\includegraphics[keepaspectratio,width=3in,height=0.75\textheight]{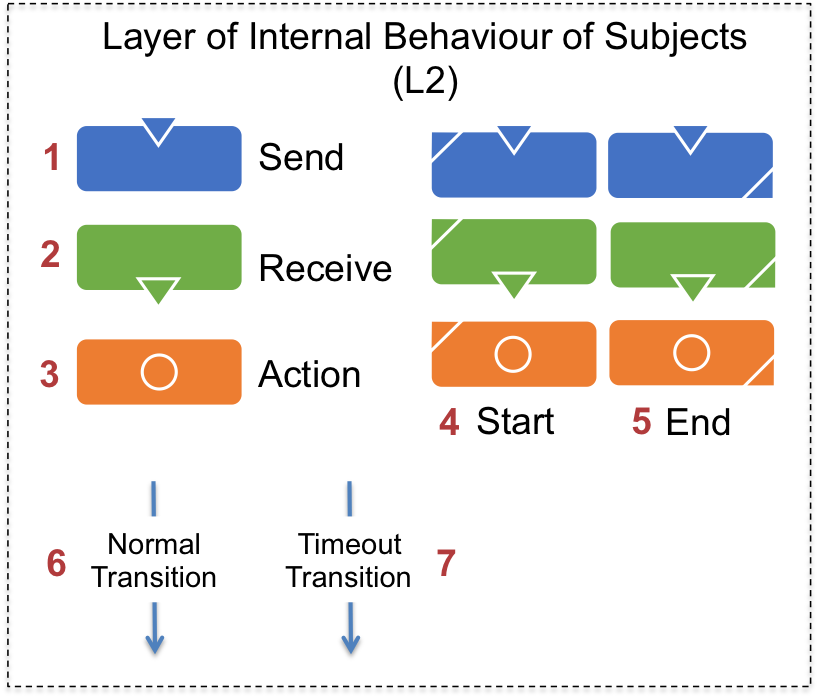}
\caption{Supported S-BPM Language Elements of the \emph{Subject Behavior Diagram} (Layer 2).}
\label{layer2}
\end{figure}

The S-BPM language~\citep{Fleischmann:2012va}~\citep{Fleischmann:2013vm}, as supported in our implementation, is depicted in \autoref{layer1} and \autoref{layer2}. The semantic meaning of these elements is the same as used in the Metasonic Suite\footnote{www.metasonic.de}, a commercial implementation of the S-BPM methodology.

The \emph{Subject Interaction Diagram} (SID) defines the \emph{Subjects} (1) and the unidirectional \emph{Channels} (4) between them. These channels establish the communication between the subjects and enable to send and receive messages at runtime. A \emph{Multi-subject} allows to send a message to more than one agent (an agent is an instance of a subject); an \emph{External-subject} allows to model a subject without knowing the internal behavior, for example another choreography participant who is not part of the own organization.

The \emph{Subject Behavior Diagram} (SBD) defines the internal behavior of a subject; \emph{Send} (1), \emph{Receive} (2) and \emph{Action} (3) are the fundamental activities for this diagram. The internal behavior of subjects has a minimum of one \emph{Start} (4) and one \emph{End} (5) activity. Any activity is marked with a flag to denote it as start or end activity. The  control flow is defined as explicit \emph{Transition} between activities. \emph{Timeout Transitions} are based on a relative time and model exceptional behavior to prevent dead lock situations or service level problems in case of no answer in a defined timeframe.

\section{Agent Based BPMS}
\label{agentbasedbpms}

A first prototype implementation of the \emph{Structured Information and Communication Technology} (StrICT) framework has been discussed in~\citep{Kotremba:2013vz}~\citep{Rass:2013kv}. This prototype had the central limit, that all subject instances (agents) had to be on the same server. This was a limitation based on the central scheduler concept; the scheduler (some software) administrates all instances of the running subject instances (please refer to ~\citep{Kotremba:2013vz}~\citep{Rass:2013kv} for a more detailed explanation). The functionality of the scheduler is similar to the \emph{Message Broker} in the \emph{ePass-IoS} architecture, an enhanced choreography implementation of S-BPM~\citep{Borgert:2011oq}.

\subsection{Architecture}
\label{architecture}

To prove the architecture concept, we have developed a fully functional application built on \emph{Microsoft Azure} cloud services. The core concepts are described in the following paragraphs and fully implement a multi-enterprise business process platform.

\begin{figure}[htbp]
\centering
\includegraphics[keepaspectratio,width=4.5in,height=0.75\textheight]{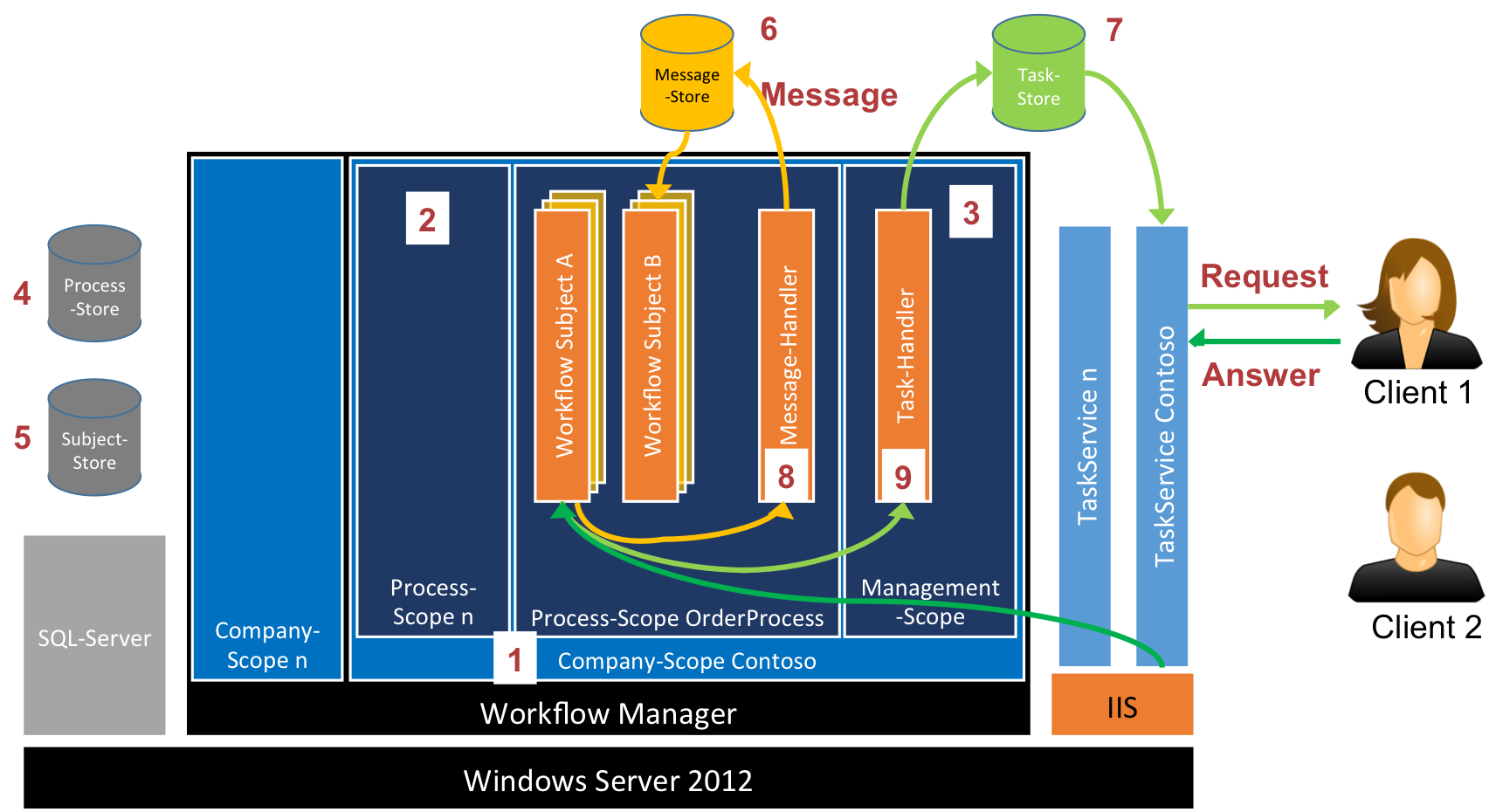}
\caption{StrICT Architecture. The processes are executed server side and the workflows are coordinated through message exchange (orange). Task requests (light green) and task answers (dark green) are routed to a client via the task service.}
\label{strict}
\end{figure}

The main work horse of our approach to realize a platform for the distributed execution of S-BPM processes is based on the functionality of \emph{Windows Workflows} as implemented in the \emph{Windows Workflow Foundation} (WF) classes, which are part of the \emph{Microsoft .NET} framework. A WF-workflow provides functionality to maintain state, get input from and send output to the outside world, provides control flow, and executes code -- this is done by so called \emph{Activities}. It can be easily seen, that it is possible to map any SID\slash SBD (that means all S-BPM constructs) on a WF-workflow. For this purpose we only need to derive classes from the WF-activity classes to implement S-BPM functionality and to match any S-BPM process onto a WF-workflow.

The complete StrICT architecture is depicted in \autoref{strict}. Processes are hosted on an instance of the \emph{Workflow Manager} (WFM), which is responsible for the hosting, administration and configuration of the subjects based on scopes, such as a \emph{Company Scope} (1), \emph{Process Scopes} (2) and a \emph{Management Scope} (3). Each company has its own \emph{Process Store} (4) and \emph{Subject Store} (5); the same for \emph{Message Store} (6) and \emph{Task Store} (7). Each company has \emph{Task Handler} (9) instances to generate new tasks and each process has \emph{Message Handler} (8) instances to manage message exchange.

\textbf{Communication between subjects} -- Messages to other subjects are routed via the internal \emph{Service} \emph{Bus} (part of \emph{Workflow} \emph{Manager}). The \emph{Message Handler} is instantiated after receiving a message and forwards it to the correct input pool (\emph{Message Store}) of the receiving subject instance; afterwards the instance is canceled. Subject instances have access to their own message pool and can choose any available message.

\textbf{User interaction} -- Interaction with process participants is done via the \emph{Task Service}. A \emph{Task} is a request to be processed by a user, typically to fill in some data into a form (or anything else). A user has full access to its list of tasks. Tasks can be routed as normal eMail to a user according the role in a S-BPM process. A task can then be answered again using a standard eMail protocol (refer to~\citep{Kotremba:2013vz} for a more enhanced explanation of such an architecture).

\textbf{External input pool} -- We have further extended this architecture by using an external service bus (\emph{Microsoft Service Bus}); this enables company to company message exchange.

\subsection{Service Interaction Pattern}
\label{serviceinteractionpattern}

To prove the functionality of our architecture we have chosen the \emph{Service Interaction Patterns}~\citep{Barros:2005tk}. They seem to be a valuable starting point, as 

\begin{quote}
They have been derived and extrapolated from insights into real-scale B2B transaction processing, use cases gathered by standardization committees (e.g. BPEL and WS-CDL), generic scenarios identified in industry standards (e.g. RosettaNet Partner Interface Protocols), and case studies reported in the literature.~\citep{Barros:2005tk}
\end{quote}

For our reference we have translated the descriptions\footnote{http:/\slash math.ut.ee\slash \ensuremath{\sim}dumas\slash ServiceInteractionPatterns} of the patterns into BPMN 2.0 collaboration diagrams, which are cross-checked with the BPMN fragments in~\citep{Weske:2012ul}. We also did a comparison on already commercially available cloud based BPMS, which we will include in our discussion. A direct execution of BPMN 2.0 models is not feasible, as the standard document defines for instance a \emph{Common Executable} subclass which does not include the elements \emph{Pool}, \emph{Lane}, \emph{Message Flow} and \emph{Data Store} needed to model collaborations.

All in ~\citep{Barros:2005tk} discussed pattern can be executed on the StrICT architecture without any restriction (for a detailed discussion of all patterns refer to~\citep{Rass:tc}).

\subsubsection{Send/Receive}

\begin{quote}
A party X engages in two causally related interactions: in the first interaction X sends a message to another party Y (the request), while in the second one X receives a message from Y (the response).
\end{quote}

\begin{figure}[htbp]
\centering
\includegraphics[keepaspectratio,width=3.5in,height=0.75\textheight]{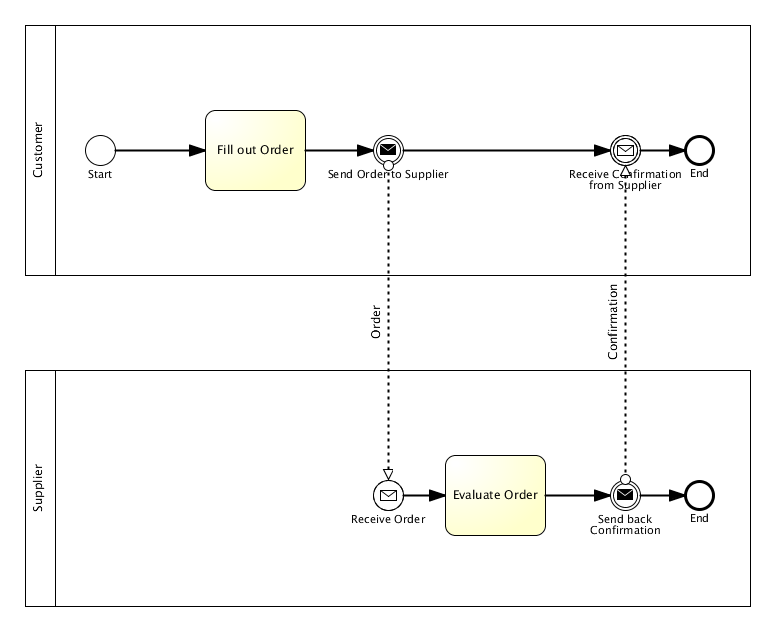}
\caption{The send/receive pattern in BPMN is fairly easy to model. In this case it consists of two pools exchanging messages.}
\label{bpmn_s_r}
\end{figure}

The BPMN process in \autoref{bpmn_s_r} shows how the pattern can look. It depicts two pools exchanging messages: A \emph{Supplier} and a \emph{Customer}. At first the customer fills out the order and sends it to the \emph{Supplier} pool. The supplier evaluates the received order and then sends back a confirmation. After the confirmation is received, the process ends. 

Of course this is just an example. It does not take into account the fact that the supplier could also not accept the order. On the other side, due to its simplicity there is no room for errors or a deadlock.

\begin{figure}[htbp]
\centering
\includegraphics[keepaspectratio,width=2.5in,height=0.75\textheight]{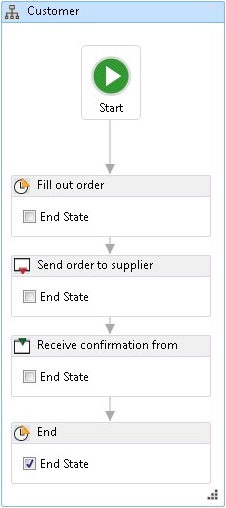}
\caption{The send/receive pattern modeled in \emph{StrICT}, showing the \emph{Customer} subject. The underlying modeling concept is plain S-BPM. The first state is a \emph{Function State}, followed by a \emph{Send}, \emph{Receive} and another \emph{Function State}. The activities only support a certain width; therefore, unfortunately, the label in the \emph{Receive State} is cut off.}
\label{promi_s_r_1}
\end{figure}

\begin{figure}[htbp]
\centering
\includegraphics[keepaspectratio,width=2.5in,height=0.75\textheight]{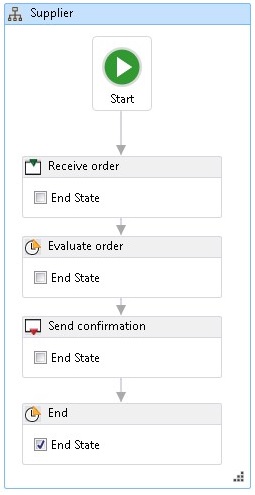}
\caption{The send/receive pattern modeled in \emph{StrICT} using S-BPM, showing the \emph{Supplier} subject.}
\label{promi_s_r_2}
\end{figure}

It is possible to model the send/receive pattern in \emph{StrICT} using plain S-BPM, as shown in \autoref{promi_s_r_1} and \autoref{promi_s_r_2}. The term "plain S-BPM" means that all elements that were used to model this process exist in the formal definition of S-BPM. Due to the concept of S-BPM being different from the other examined approaches, there are two figures needed to show the whole process. Each figure shows one subject taking part in the process. The process is started by the \emph{Customer} subject filling out an order and sending it to the other subject, the \emph{Supplier}. The \emph{Supplier} subject (\autoref{promi_s_r_2}) receives the order, evaluates it and then sends back a confirmation message. After that, the process ends for the supplier. The customer then receives the confirmation and is also done with the process.

This is a very simple S-BPM process with little room for error. However, there are some visual impairments in the figure, e.g. when the label of the states is cut off. But since the main objective was to achieve functionality over usability, this is a necessary evil.

\subsubsection{Racing incoming messages}

\begin{quote}
A party expects to receive one among a set of messages. These messages may be structurally different (i.e. different types) and may come from different categories of partners. The way a message is processed depends on its type and/or the category of partner from which it comes.
\end{quote}

\begin{figure}[htbp]
\centering
\includegraphics[keepaspectratio,width=3.5in,height=0.75\textheight]{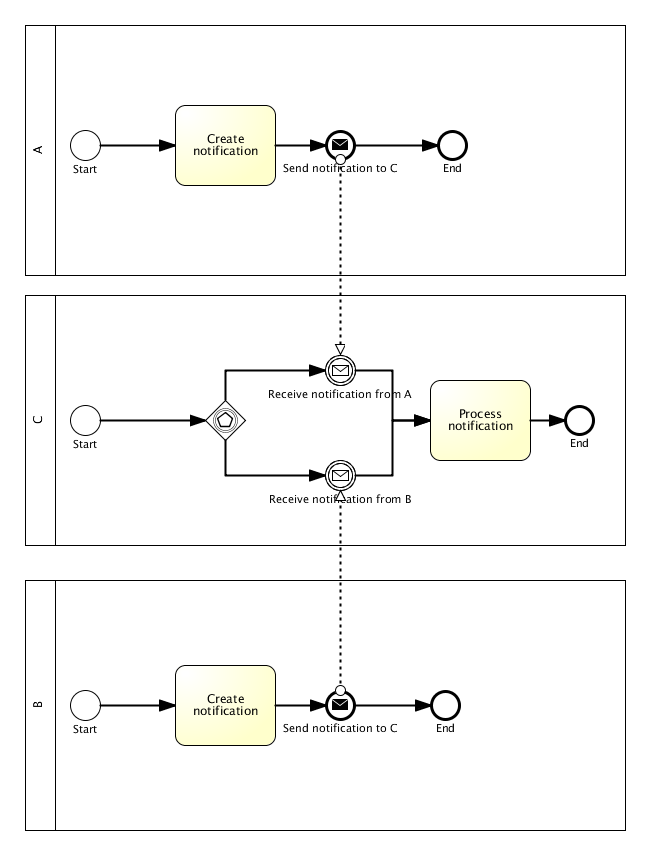}
\caption{Racing incoming messages: To prove the concept, the pools were named alphabetically: \emph{A, B} and \emph{C}. The first two send a message to the latter one, the first arriving message is accepted and the other one is discarded.}
\label{bpmn_r}
\end{figure}

BPMN is able to model this pattern, as shown in \autoref{bpmn_r}. The figure shows an example process between three alphabetically labeled pools. All three pools start at the same time. \emph{A} and \emph{C} create a notification and send it to \emph{C}. The first arriving message triggers the \emph{Event Based Gateway}, making the process continue. The second message is discarded.

This sample process is deliberately kept simple. The only theoretical but small possibility of an error occurs when the notifications of \emph{A} and \emph{B} arrive at exactly the same time. But due to the tasks being user tasks, this possibility is almost nonexistent. Also, the possibility of one of the two sending pools to not send a message was not included. This could lead to a deadlock situation if both pools choose to not send a message.

\begin{figure}[htbp]
\centering
\includegraphics[keepaspectratio,width=2.5in,height=0.75\textheight]{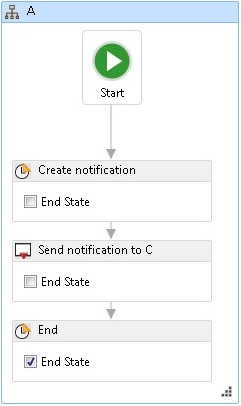}
\caption{Racing incoming messages: This figure shows subject \emph{A} participating in the incoming message race. The behavior is simple and plain S-BPM.}
\label{promi_r_a}
\end{figure}

\begin{figure}[htbp]
\centering
\includegraphics[keepaspectratio,width=2.5in,height=0.75\textheight]{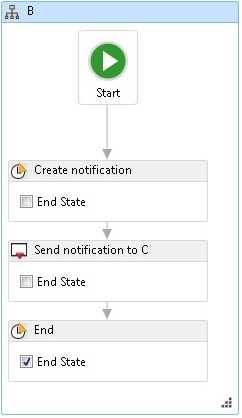}
\caption{Racing incoming messages: Subject \emph{B} has the same behavior as subject \emph{A}.}
\label{promi_r_b}
\end{figure}

\begin{figure}[htbp]
\centering
\includegraphics[keepaspectratio,width=2.5in,height=0.75\textheight]{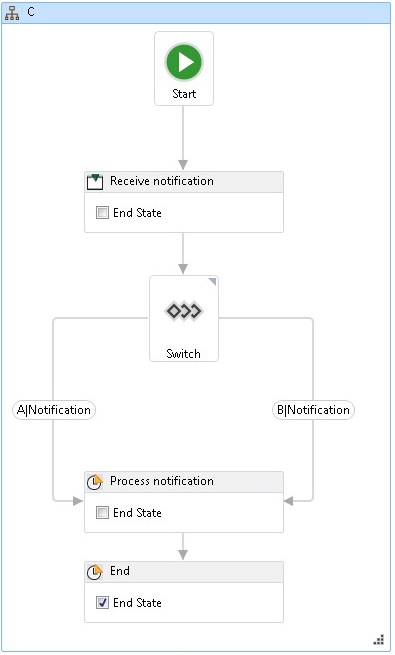}
\caption{Racing incoming messages: Subject \emph{C} decides which message is accepted and, therefore, decides who wins the message race.}
\label{promi_r_c}
\end{figure}

This pattern can also be modeled using \emph{StrICT} and plain S-BPM. \autoref{promi_r_a}, \autoref{promi_r_b} and \autoref{promi_r_c}) show the three participating subjects, named alphabetically. Subject \emph{A} and \emph{B} have the same behavior. Both try to send a message to the third subject, \emph{C}. Both create a notification at first and then send it to \emph{C}. 

\emph{C} receives said notification. In the S-BPM concept, the recipient must actively accept a message to receive it. So in that case \emph{C} must choose which message to receive. In the best-case scenario, only one message is present. But in the worst case scenario, both have already arrived when \emph{C} wants to receive his messages. In that case \emph{C} sees by the timestamp which message arrived first and must make the right choice -- this is not done automatically. The concept here relies on the participation of the user. Theoretically, \emph{C} could also choose the message which arrived last. No matter what, the second message is discarded afterwards and the process ends. 

\subsubsection{One-to-many send/receive}

\begin{quote}
A party sends a request to several other parties, which may all be identical or logically related. Responses are expected within a given timeframe. However, some responses may not arrive within the timeframe and some parties may even not respond at all. The interaction may complete successfully or not depending on the set of responses gathered. 
\end{quote}

\begin{figure}[htbp]
\centering
\includegraphics[keepaspectratio,width=3.5in,height=0.75\textheight]{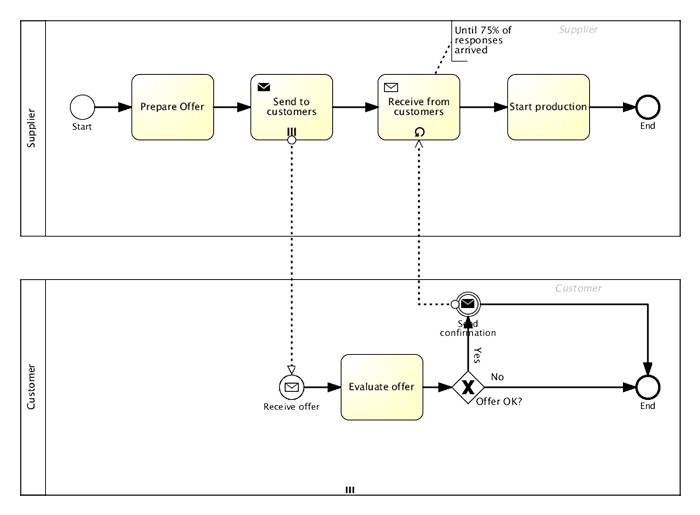}
\caption{One-to-many send/receive: This figure shows the \emph{Customer} and \emph{Supplier} pools exchanging messages. The two special features about this process are the multi-instance tasks, the multi-instance pool and the loop activity.}
\label{bpmn_otm_r}
\end{figure}

\autoref{bpmn_otm_r} proves that this pattern can be modeled in BPMN. The tricky part with this pattern is that the amount of "other parties" to which a request is sent, is not known at the time of modeling. The process must be able to provide a varying number of recipients. Fortunately this is possible in BPMN by using multi-instance concepts in modeling. The process is started by the supplier, preparing an offer. The offer is then sent to a certain number of customers specified at runtime. The \emph{Send to customers} task is executed until all offers are sent, marked by the three vertical lines shown in the task. Each offer is sent to the \emph{Customer} pool and starts a new instance of said pool. Normally, only one instance of a pool can be active at the same time. But this is a multi-instance pool, marked also by the three vertical lines on the bottom of the pool. All active customer instances then evaluate the offer. If the offer is not OK, the process ends right away for that instance. Otherwise a confirmation is sent back and the process ends. The supplier then receives the confirmations until at least 75\% have arrived. 

Even though this process only needs a few symbols to model, the underlying concept of multiple instances is very complicated. But it complies to BPMN standards and is a valid process. In a real world process, the \emph{Supplier} pool would be accompanied by a timeout (or similar) to prevent a deadlock.

\begin{figure}[htbp]
\centering
\includegraphics[keepaspectratio,width=3.0in,height=0.75\textheight]{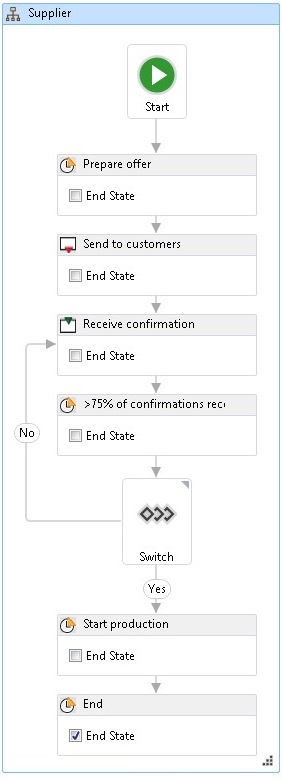}
\caption{One-to-many send/receive: The \emph{Supplier} subject initiates the process and sends out the notifications. Each answer must be actively received. There is no automation mechanism for the amount of received messages.}
\label{promi_o_t_m_s}
\end{figure}

\begin{figure}[htbp]
\centering
\includegraphics[keepaspectratio,width=3.0in,height=0.75\textheight]{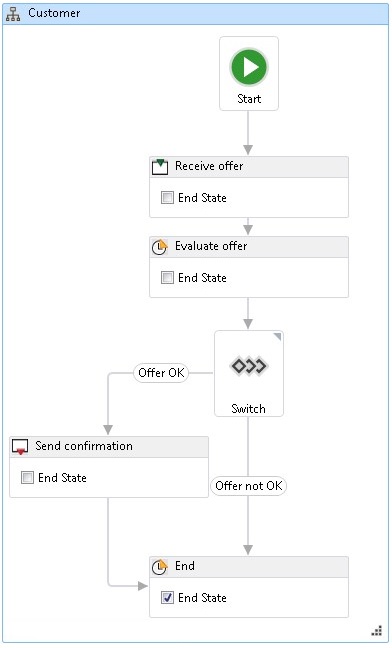}
\caption{One-to-many send/receive: The customer is a multi subject with a simple behavior -- it just decides whether to send back a confirmation or not.}
\label{promi_o_t_m_c}
\end{figure}

\autoref{promi_o_t_m_s} and \autoref{promi_o_t_m_c} show the two subjects participating in this pattern, effectively proving it to be possible using \emph{StrICT} with plain S-BPM. \autoref{promi_o_t_m_s} shows the \emph{Supplier} subject which starts the process. The supplier prepares and sends an offer to the customers. The customers, in this case, are a multi-subject, which means that the \emph{Customer} subject can be instantiated multiple times. There is no special visual aid to mark this except for the description of the state. In the send state, multiple offers are sent to the customer, instantiating the subject several times. The exact amount is not known at the time of modeling and is decided during runtime. After sending, the \emph{Customer} subject waits for answers. 

Each instantiated customer is an agent, which is the S-BPM term for an instantiated subject. Each agent apart from the \emph{Supplier} is actively receiving the offer and evaluating it. If the offer is OK, a confirmation is sent back. Otherwise, nothing happens. The process is finished for each of these agents in either case.

The \emph{Supplier} can then actively receive the arrived messages but must do so one at a time. After each message the agent must check if more than 75\% of the messages have arrived. If this is the case, production begins and the process ends. Again, this is not an automated task and relies on the user. The user must actively check the amount of answers received and go back to receiving more confirmations if necessary.

This process contains the possibility of a deadlock when less than 75\% of the customers send back a confirmation.

\subsubsection{Multi-responses}

\begin{quote}A party X sends a request to another party Y. Subsequently, X receives any number of responses from Y until no further responses are required. The trigger of no further responses can arise from a temporal condition or message content, and can arise from either X or Y's side. Responses are no longer expected from Y after one or a combination of the following events: (i) X sends a notification to stop; (ii) a relative or absolute deadline indicated by X; (iii) an interval of inactivity during which X does not receive any response from Y; (iv) a message from Y indicating to X that no further responses will follow. From this point on, no further messages from Y will be accepted by X.
\end{quote}

\begin{figure}[htbp]
\centering
\includegraphics[keepaspectratio,width=3.5in,height=0.75\textheight]{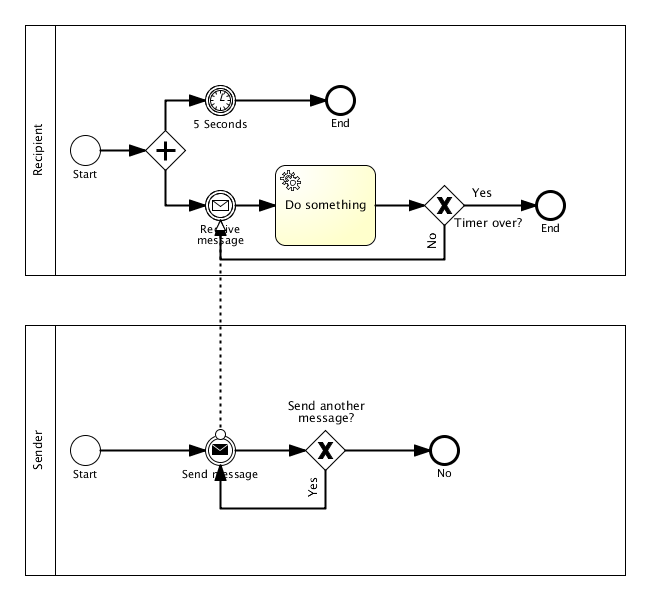}
\caption{Multi-responses: In this instance the only task is not a user task but a service task. Service tasks are automated tasks.}
\label{bpmn_m_r}
\end{figure}

With BPMN it is also possible to model this pattern, as seen in \autoref{bpmn_m_r}. The process consists, again, of two pools exchanging a message. The concept is that one of the two pools is able to send messages until it decides to stop. The process starts for both pools simultaneously. The recipient starts a timer and waits for the first message while the sender starts and sends the first message. After the message is received, the recipient checks the timer via a service task (meaning that this could be an automated task) and only goes back to receiving another message if the timer is still running. The sender then can decide whether to send another message or not. After five seconds the process is over. The process is then ended. If, for some reason, the process does not end after the timer runs out, there is a failsafe in place: if the timer is checked by the service task and it is determined that the timer ran out, the process ends.

The process could, of course, also continue after all messages arrived. But in this case the real world scenario could be that the messages are stored in a database and no further action is needed from this process. The tricky situation is to not let the part where the message is received, slip into a deadlock. If not carefully modeled, this could happen if the recipient decides to receive another message but the sender simultaneously decides not to send anymore messages. 

\begin{figure}[htbp]
\centering
\includegraphics[keepaspectratio,width=3.0in,height=0.75\textheight]{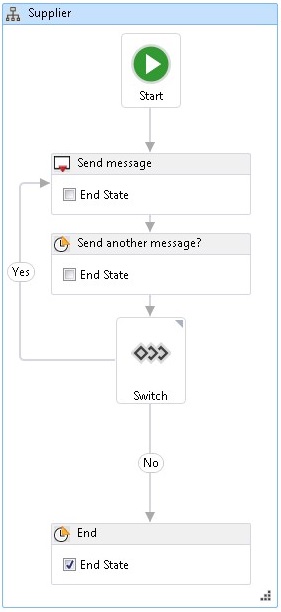}
\caption{Multi-responses: The \emph{Supplier} subject is still S-BPM and uses a simple pattern of sending one specific message all over again.}
\label{promi_m_r_s}
\end{figure}

\begin{figure}[htbp]
\centering
\includegraphics[keepaspectratio,width=3.0in,height=0.75\textheight]{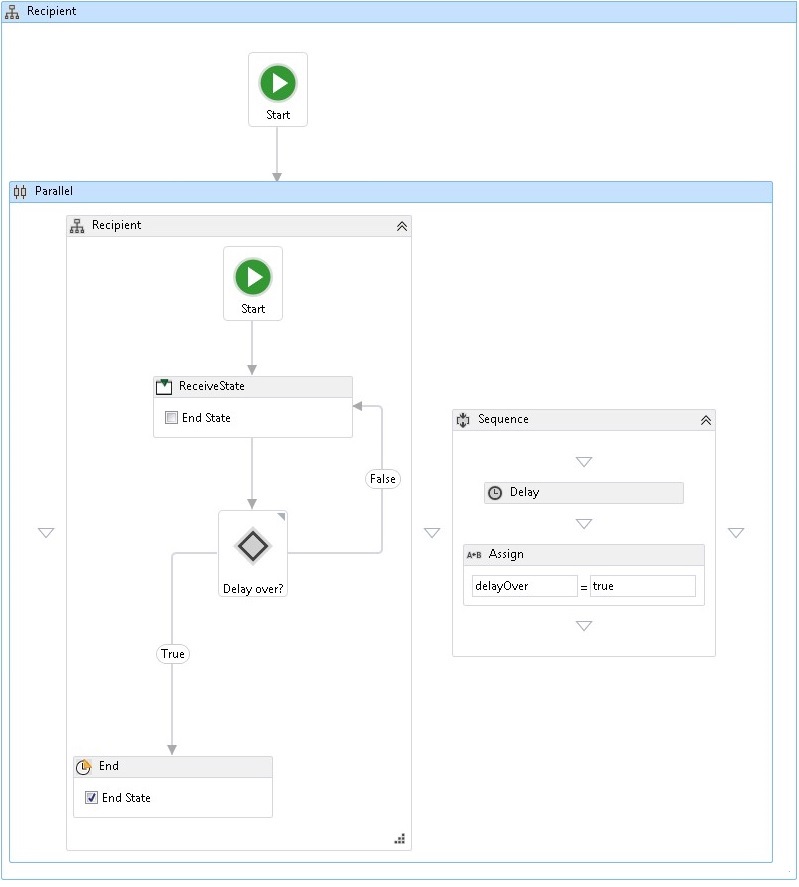}
\caption{Multi-responses: The \emph{Recipient} subject contains out-of-the-box \emph{Windows Workflow} technology, namely the \emph{Parallel} activity, the \emph{Delay} activity and the \emph{Assign} activity. Together they make the pattern possible, effectively and automatically stopping the \emph{Recipient} from receiving more messages when the time is up.}
\label{promi_m_r_r}
\end{figure}

This process was also modeled using two subjects which are depicted in \autoref{promi_m_r_s} and \autoref{promi_m_r_r}. 

The process starts with the \emph{Supplier} subject starting to send messages to the \emph{Recipient} subject. After each message, the supplier has the possibility of sending another message or to end the process. 

The \emph{Recipient} enters a \emph{Parallel} state at runtime, which is an out-of-the-box \emph{Windows Workflow} activity. The \emph{Parallel} activity ensures that the recipient can receive messages only until the delay is over. If the delay is over the process ends. 

If the \emph{Supplier} decides to continue sending messages after the time is over, it can do so. In S-BPM the subjects are decentralized from each other. But the messages will have no effect. There is the theoretical possibility of a deadlock if the \emph{Supplier} subject does not stop to send messages. 

\subsubsection{Contingent Request}

\begin{quote}
A party X makes a request to another party Y. If X does not receive a response within a certain timeframe, X alternatively sends a request to another party Z, and so on.
\end{quote}

\begin{figure}[htbp]
\centering
\includegraphics[keepaspectratio,width=3.5in,height=0.75\textheight]{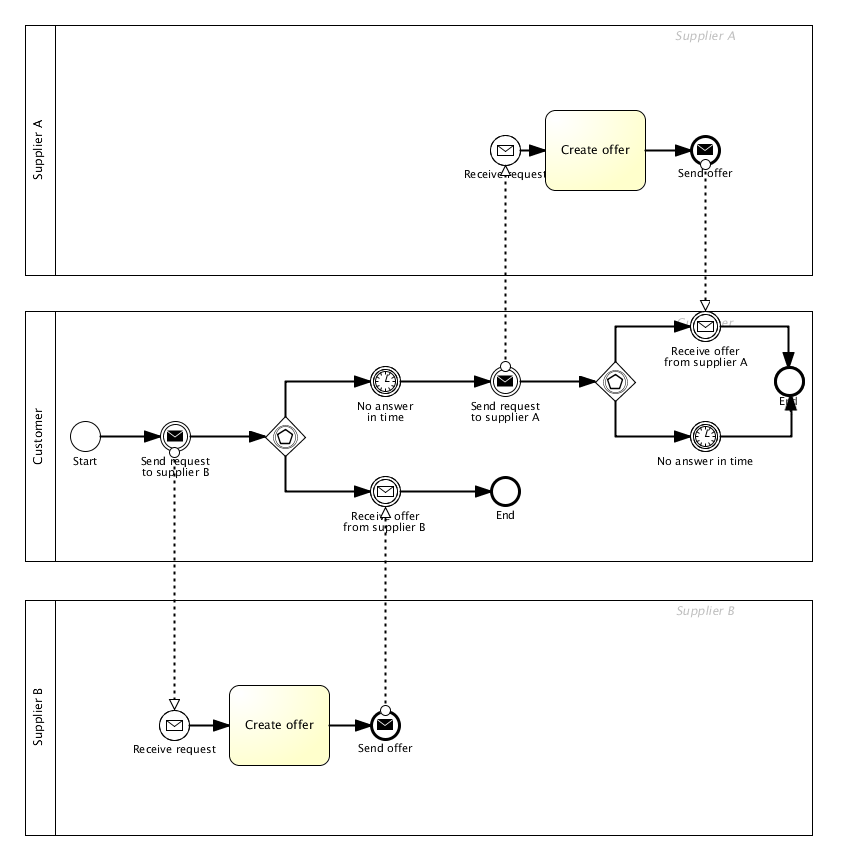}
\caption{The contingent request pattern modeled in BPMN 2.0.}
\label{bpmn_c_r}
\end{figure}

This pattern modeled with BPMN can be seen in \autoref{bpmn_c_r}. It consists of two suppliers and one customer. The process starts with the customer sending a request to a supplier and simultaneously starting a timer. The \emph{Supplier B} pool could now receive the request, create an offer and send it back. In this case the process ends right away. If that is not the case, the process continues when the timer is triggered. After that, the same request is sent to the other supplier and a message from the first pool is not accepted anymore. The procedure here is the same: either the \emph{Supplier A} pool sends back an offer or the timer is triggered. Either way, the process is ended right after. It could be the case that one or both of the suppliers never get beyond the \emph{Create offer} task, placing that specific pool in a deadlock. 

This process could also be modeled using only one multi-instance pool for the \emph{Supplier}. But the process was modeled in a static way to focus on the understanding of the reader. A dynamic process would feature one multi-instance \emph{Supplier} pool and a recursion in the \emph{Customer} pool, enabling the customer to try as many alternative suppliers as needed.

\begin{figure}[htbp]
\centering
\includegraphics[keepaspectratio,width=3.0in,height=0.75\textheight]{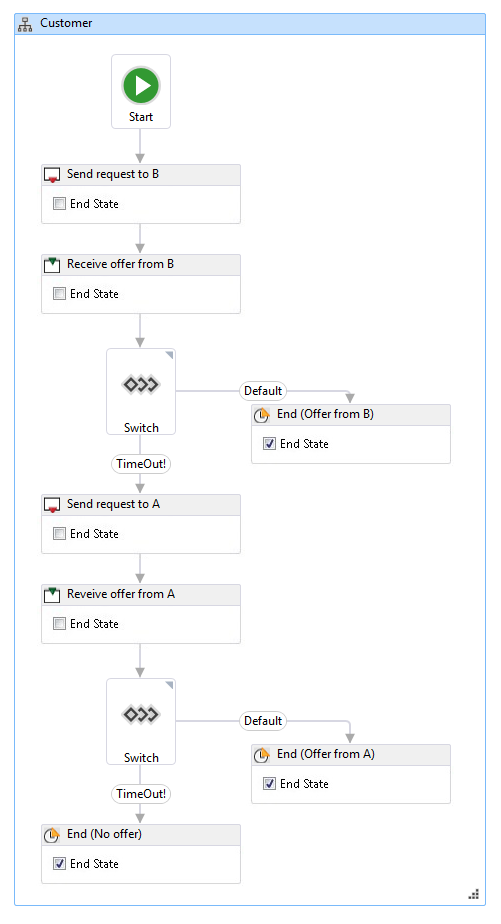}
\caption{The contingent request pattern modeled in StrICT WF-workflows (Customer)}
\label{promi_c_r_c}
\end{figure}

\begin{figure}[htbp]
\centering
\includegraphics[keepaspectratio,width=\textwidth,height=3.0in]{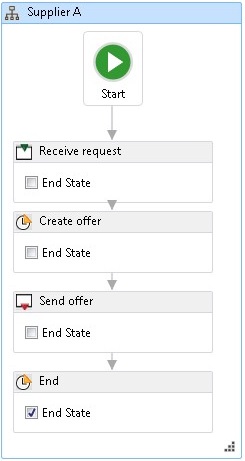}
\caption{The contingent request pattern modeled in StrICT WF-workflows (Supplier); we have modeled explicitly both subjects instead of one multi-subject for clarity and conformance with our BPMN model}
\label{promi_c_r_ab}
\end{figure}

The process to prove the pattern consists of three subjects (or one subject and a multi-subject); the WF-workflow models are depicted in \autoref{promi_c_r_c} and \autoref{promi_c_r_ab} respectively.

The process starts with the customer sending a request to the \emph{Supplier B} subject. After that, the \emph{Windows Workflow} activity simultaneously waits for the offer to be received and starts a timer. If the offer from \emph{Supplier B} is received in time, the process ends. 

Else, a request is sent to the other supplier, \emph{Supplier A}. A similar pattern happens: the \emph{Customer} subject simultaneously waits for a message to arrive and a timer to run out. If the message arrives before the delay is triggered, the process ends. If the timer runs out before the offer arrives, the same thing happens. No deadlock can happen, even if the suppliers decide not to play along because there are timeouts in place for both \emph{Receive States}.

\subsubsection{Atomic Multicast Notification}

\begin{quote}
A party sends notifications to several parties such that a certain number of parties are required to accept the notification within a certain timeframe. For example, all parties or just one party are required to accept the notification. In general, the constraint for successful notification applies over a range between a minimum and maximum number.
\end{quote}

\begin{figure}[htbp]
\centering
\includegraphics[keepaspectratio,width=4.5in,height=0.75\textheight]{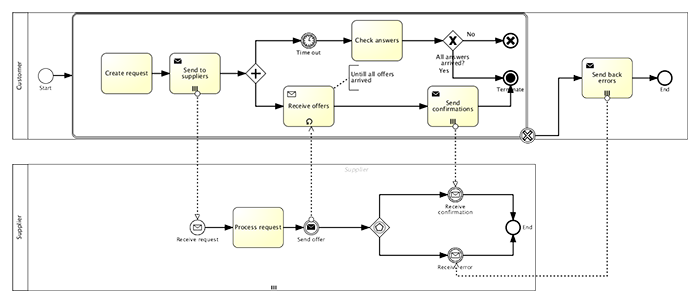}
\caption{The atomic multicast also requires the use of multiple instances. For the concept of atomic transactions, the use of the BPMN \emph{Transaction Sub-Process, Intermediate Cancel Event}, and \emph{Cancel End Event} is necessary.}
\label{bpmn_a_m}
\end{figure}

Even though it requires some time, the atomic multicast is possible in BPMN as \autoref{bpmn_a_m} shows. The concept of a transaction is present in the BPMN 2.0 standard. The process itself consists of the already known \emph{Supplier} and \emph{Customer} pools. The supplier, in this case, is a multi-instance pool. Like in one of the previous processes, the amount of outgoing messages is dynamic and not known at the time of modeling. The process starts with the customer creating a request. This request is then sent to various suppliers, using a multi-instance send task. The concept of an atomic transaction is, that it is only successful if everything goes fine, otherwise there is an error. This is done by a multi-instance receive task, which is only triggered if an answer to every single request has arrived. If this is the case, a confirmation is sent to every instance of the \emph{Supplier} pool and the process is then terminated. In parallel, there is also a timer. If the timer is triggered, the amount of arrived messages is checked. If even a single one does not arrive, error messages are sent to all suppliers and the process is terminated.

The \emph{Supplier} is modeled in a simple way. The request is received and processed, then an offer is sent. Then either a confirmation or an error message arrives and the process is finished for the pool.

\begin{figure}[htbp]
\centering
\includegraphics[keepaspectratio,width=3.0in,height=0.75\textheight]{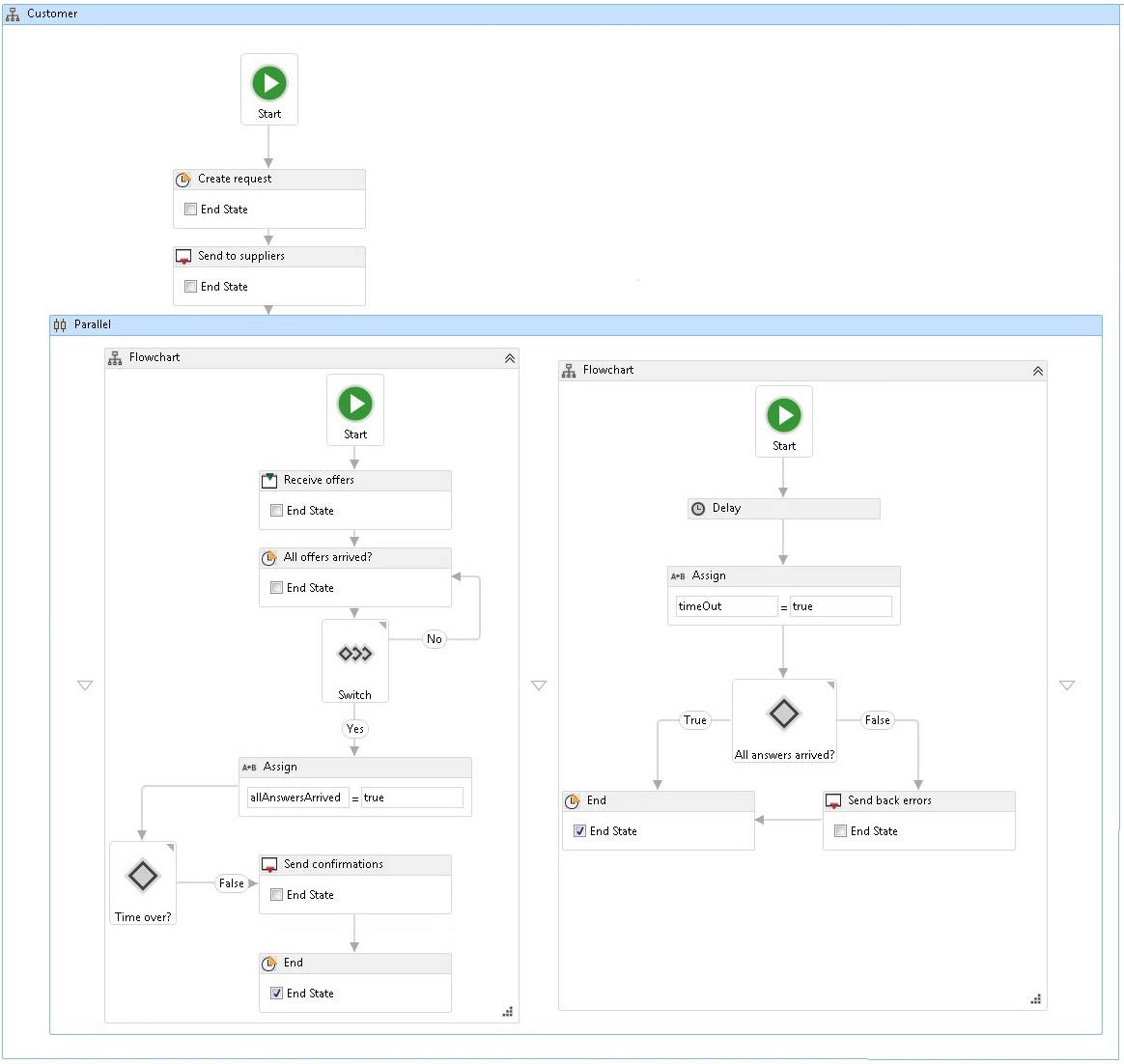}
\caption{Atomic Multicast Notification: The \emph{Customer} subject is ready to receive messages within a certain timeframe. Due to the atomic concept, it sends out error messages even if one offer did not arrive.}
\label{promi_a_m_c}
\end{figure}

\begin{figure}[htbp]
\centering
\includegraphics[keepaspectratio,width=3.0in,height=0.75\textheight]{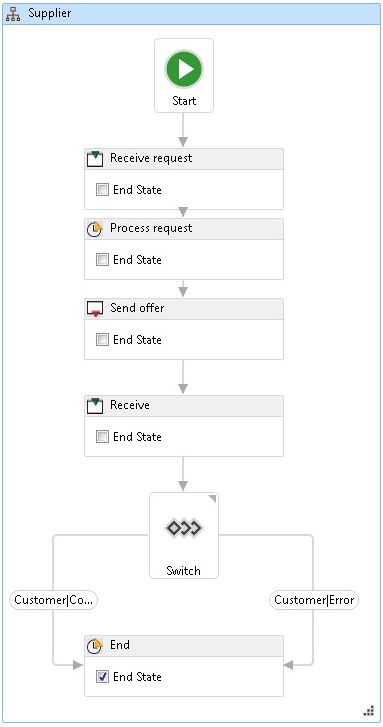}
\caption{Atomic Multicast Notification: After receiving a request, the \emph{Supplier} multi-subject simply processes the request and sends an offer. Depending on the arriving answer, the process ends either way. Due to limitations of \emph{Windows Workflow} the name of the arriving confirmation message is not fully displayed.}
\label{promi_a_m_s}
\end{figure}

The atomic multicast can be modeled using \emph{StrICT} and two subjects. The S-BPM notation was enhanced with \emph{Windows Workflow}, more specifically with an \emph{Assign} activity, to reduce the amount of required user interactions by one. \autoref{promi_a_m_c} shows the \emph{Customer} subject while \autoref{promi_a_m_s} focuses on the \emph{Supplier} subject, which is a multi-instance subject.

The customer starts the process by creating a request and sending it to multiple suppliers. After that, the customer simultaneously waits for any arriving offers. After each arriving offer the customer checks if all offers have arrived. If that is not the case, the process goes back to the \emph{Receive State}. Otherwise a variable is assigned to be \emph{true}. If the time runs out while this variable is being assigned, nothing happens. Only if all offers arrive within the timeframe, confirmations will be sent to all participating suppliers, and then the process ends.

If the time runs out with even one offer not arriving, error messages are sent and the process also ends. 

Compared to the \emph{Customer} subject, the \emph{Supplier} multi-subject is fairly easy. It receives a request, processes it, sends an offer back and waits for the answer. The answer is either an error or a confirmation. No matter what, the process still ends. 

\subsubsection{Request with Referral}

\begin{quote}
Party A sends a request to party B indicating that any follow-up response should be sent to a number of other parties (P1, P2, ..., Pn) depending on the evaluation of certain conditions. While faults are sent by default to these parties, they could alternatively be sent to another nominated party (which may be party A).
\end{quote}

\begin{figure}[htbp]
\centering
\includegraphics[keepaspectratio,width=3.0in,height=0.75\textheight]{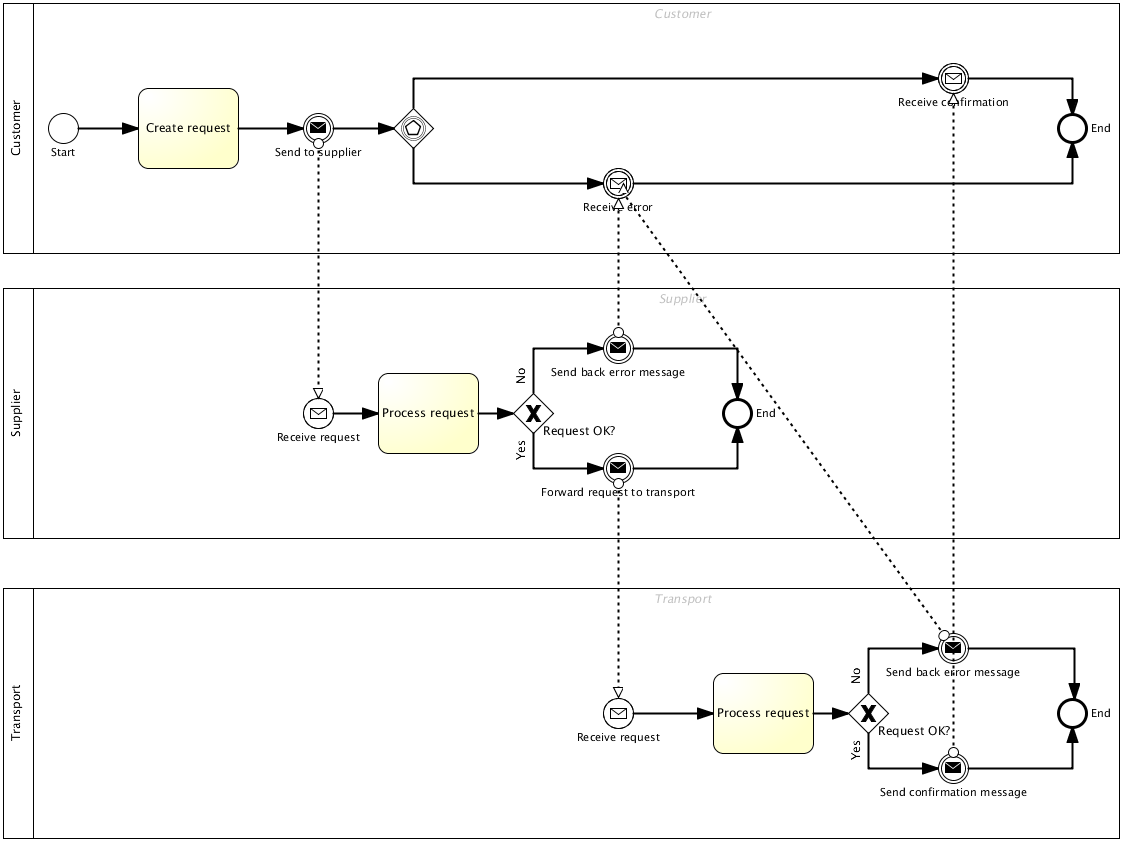}
\caption{Request with Referral: This process consists of three pools exchanging messages. The \emph{Supplier} pool only acts as a proxy to forward the original request to the \emph{Transport} pool which, in the desired case, then communicates with the \emph{Customer} pool.}
\label{bpmn_r_w_r}
\end{figure}

The first of the routing patterns is perfectly possible in BPMN, as shown in \autoref{bpmn_r_w_r}. The process consists of three pools: the \emph{Customer, Supplier} and \emph{Transport} pool. The underlying concept is that the supplier acts as a proxy and can not send a confirmation message on its own. The process starts with the customer creating a request and sending it to the supplier. The supplier processes the request. If it is not okay then the supplier sends back an error message right away and the process is finished for all participating pools. Otherwise the request is forwarded to the \emph{Transport} pool. The behaviour here is similar: After processing the request, an error message is sent to the customer if it is not okay. Otherwise, a confirmation message is sent back to the customer and the process ends. 

There is no conceptual way for a deadlock, therefore no \emph{Termination} event is needed. The only way a deadlock can occur is if one of the pools decides not to participate in the process.

\begin{figure}[htbp]
\centering
\includegraphics[keepaspectratio,width=3.0in,height=0.75\textheight]{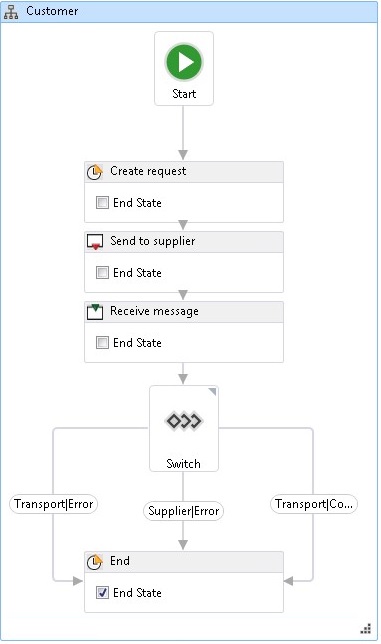}
\caption{Request with Referral: The \emph{Customer} subject can receive one out of three possible messages in return. Unfortunately, the confirmation message is not fully displayed.}
\label{promi_r_w_r_c}
\end{figure}

\begin{figure}[htbp]
\centering
\includegraphics[keepaspectratio,width=3.0in,height=0.75\textheight]{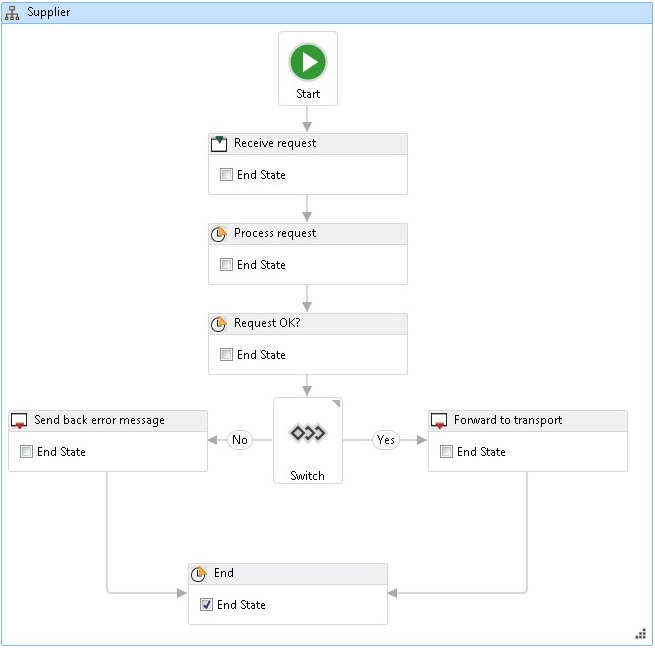}
\caption{Request with Referral: After receiving a request, the supplier either sends back an error message or forwards the message to the \emph{Transport} subject. After that, the subject is done with the process.}
\label{promi_r_w_r_s}
\end{figure}

\begin{figure}[htbp]
\centering
\includegraphics[keepaspectratio,width=3.0in,height=0.75\textheight]{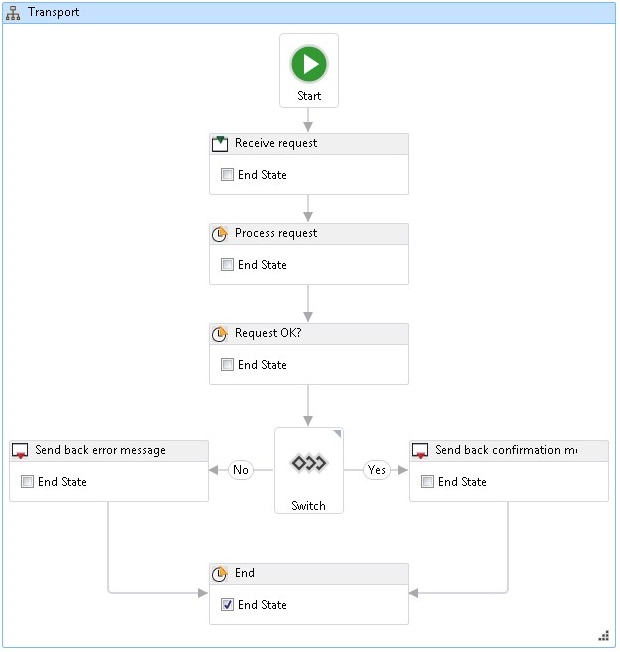}
\caption{Request with Referral: The behavior of the \emph{Transport} subject is similar to the \emph{Supplier} subject, except that it does not forward the request but sends a confirmation message back to the customer.}
\label{promi_r_w_r_t}
\end{figure}

The first of the routing patterns can be modeled using \emph{StrICT} and plain S-BPM without any enhancements. \autoref{promi_r_w_r_c}, \autoref{promi_r_w_r_s} and \autoref{promi_r_w_r_t} show the participants in the process, namely the \emph{Customer}, the \emph{Supplier} and the \emph{Transport} subject.

The process starts with the customer creating a report and sending it to the supplier. The supplier receives the request and processes it. If the request is not OK an error message is sent back to the customer and the process ends right away for both subjects. Otherwise, the request is forwarded to the \emph{Transport} subject. 

After the \emph{Transport} subject receives the message, it also processes it and decides whether it is OK or not. If it is not OK an error message is sent back and the process ends for all three subjects. Otherwise a confirmation message is sent to the customer and the process also ends for all three subjects.

\subsubsection{Relayed Request}

\begin{quote}
Party A makes a request to party B which delegates the request to other parties (P1, ..., Pn). Parties P1, ..., Pn then continue interactions with party A while party B observes a "view" of the interactions including faults. The interacting parties are aware of this "view" (as part of the condition to interact).
\end{quote}

\begin{figure}[htbp]
\centering
\includegraphics[keepaspectratio,width=3.0in,height=0.75\textheight]{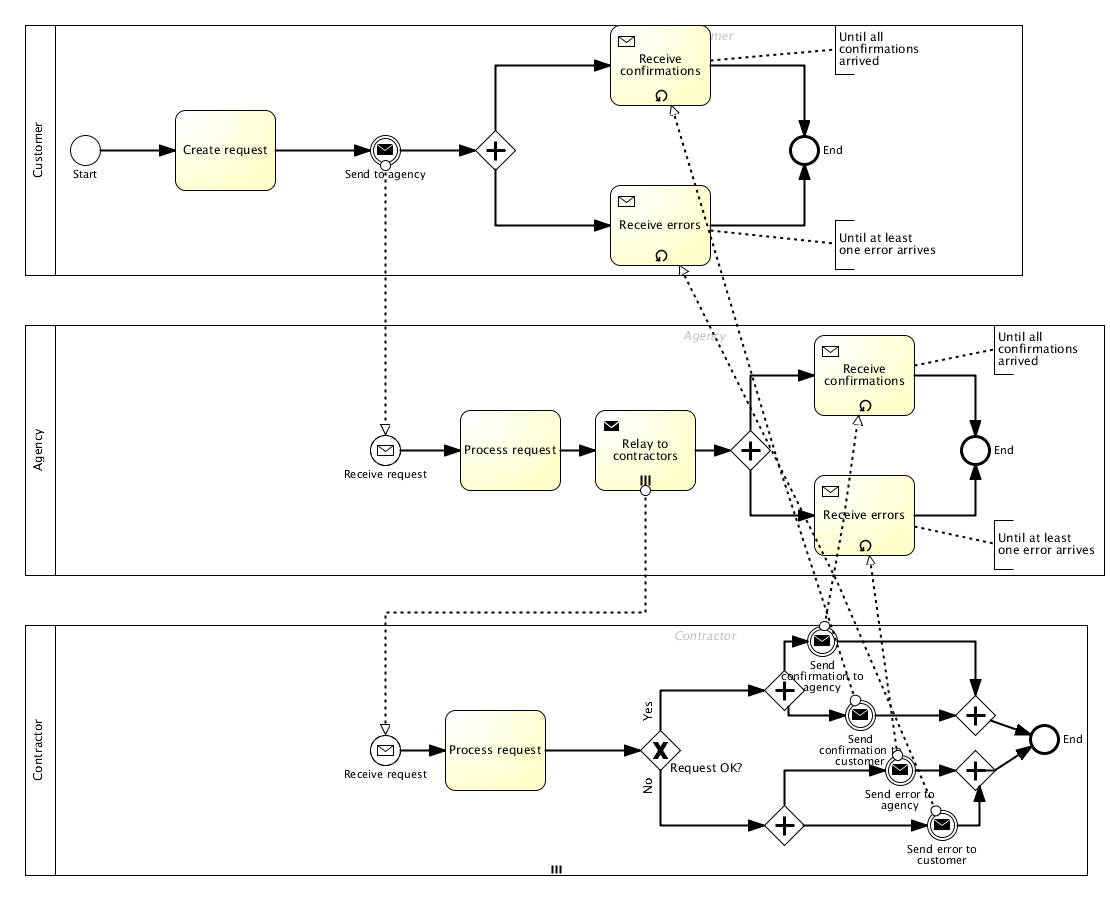}
\caption{Relayed Request: These three pools, of which one is a multi-instance pool, exchange messages to create the relayed request pattern in BPMN. The overlapping message flows can look confusing, but looking at the source should clarify the destination.}
\label{bpmn_r_r}
\end{figure}

The relayed request pattern process model can be confusing to look at, as seen in \autoref{bpmn_r_r}. Unfortunately, several message flows overlap and can be difficult to follow. Still, it is a valid process model. The process consists of three pools, namely \emph{Customer, Agency}, and \emph{Contractor}. The latter one is a multi-instance pool because at the time of modeling the amount of contractors is not known.

The process starts with the customer creating a request and sending it to an agency. The agency then receives the request, processes it, and then relays it to a varying amount of contractors. The individual contractors then receive and process the request. If the request is okay, a confirmation is sent back to both the agency and the customer. Otherwise an error message is sent back to both pools. After doing so, the process ends for each individual customer.

For the customer and the agency, the rest of the process remains equal; for both, the process only continues if either all confirmations or at least one error arrive. Either way, the process ends.

Even though it requires a lot of message flows, the underlying concept is simple and deadlock proof.

\begin{figure}[htbp]
\centering
\includegraphics[keepaspectratio,width=3.0in,height=0.75\textheight]{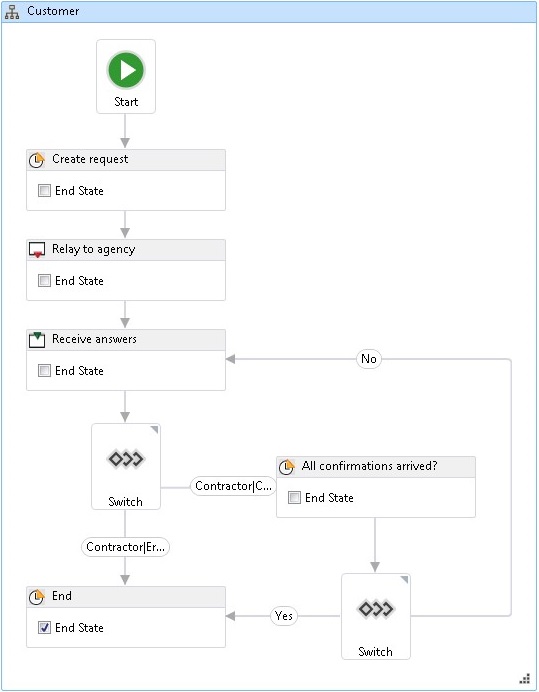}
\caption{Relayed Request: The \emph{Customer} is modeled straightforward as a sending and receiving subject.}
\label{promi_r_r_c}
\end{figure}

\begin{figure}[htbp]
\centering
\includegraphics[keepaspectratio,width=3.0in,height=0.75\textheight]{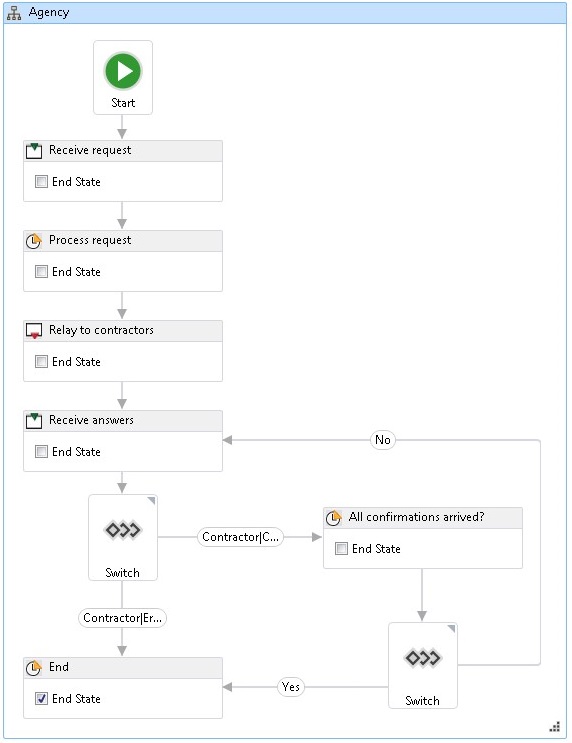}
\caption{Relayed Request: The \emph{Agency} shares most of the behavior with the \emph{Customer} subject and reacts in the same way to the received messages from the \emph{Contractors}. The point of the pattern is the relayed request from the \emph{Agency} to the \emph{Contractor} multi subject.}
\label{promi_r_r_a}
\end{figure}

\begin{figure}[htbp]
\centering
\includegraphics[keepaspectratio,width=3.0in,height=0.75\textheight]{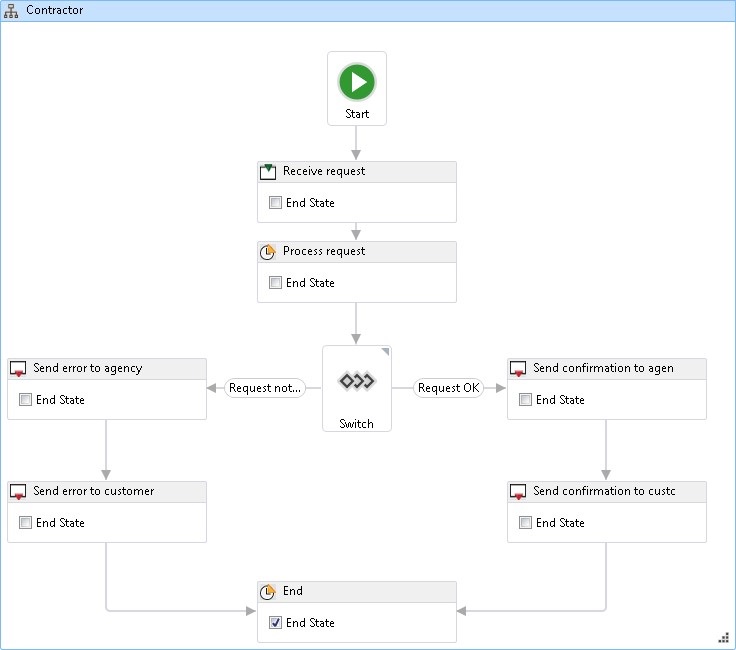}
\caption{Relayed Request: The \emph{Contractor} is a multi-subject used to send either error or confirmation messages. To stay with plain S-BPM, the messages are sent sequentially to the agency and the customer. The messages could also be sent to the respective subjects at the same time using a \emph{Parallel} activity from \emph{Windows Workflow}.}
\label{promi_r_r_c2}
\end{figure}

The relayed request can also be modeled using \emph{StrICT} and plain S-BPM. The three participating subjects are shown in \autoref{promi_r_r_c}, \autoref{promi_r_r_a} and \autoref{promi_r_r_c2}. The last one, the \emph{Contractor} subject, is a multi subject.

The customer starts the whole process by creating a request and relaying it to the agency. After that, the customer waits for answers. 

After receiving the request, the agency processes it and relays it to a number of independent contractors. After that, the behavior equals the behavior of the \emph{Customer} subject.

Since the \emph{Contractor} is a multi-subject, several instances are actively receiving their respective requests. After processing it, they decide to either approve or not approve the request. If the request is OK, confirmations are sent to the agency and the customer. Otherwise both subjects receive error messages. 

For this pattern the \emph{Customer} and \emph{Agency} subjects react in the same way. When receiving a confirmation message, both check if all confirmations have arrived. If that is the case the process ends. Otherwise more messages are received. A single received error message is enough for the process to end because it can only be successful if all confirmation and no error messages arrived.

\subsubsection{Dynamic Routing}

\begin{quote}
A request is required to be routed to several parties based on a routing condition. The routing order is flexible and more than one party can be activated to receive a request. When the parties that were issued the request have completed, the next set of parties are passed the request. Routing can be subject to dynamic conditions based on data contained in the original request or obtained in one of the 'intermediate steps'.
\end{quote}

\begin{figure}[htbp]
\centering
\includegraphics[keepaspectratio,width=3.0in,height=0.75\textheight]{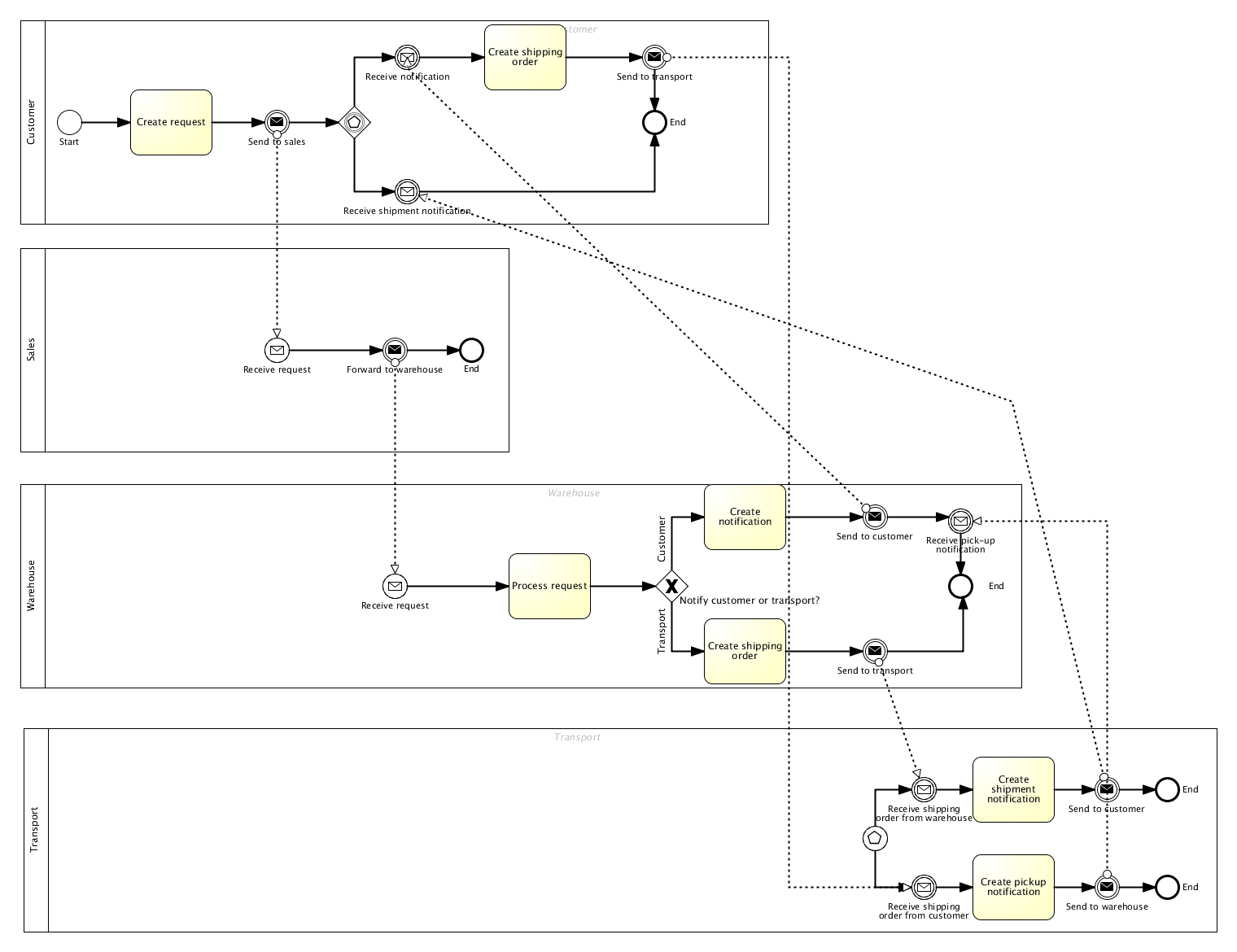}
\caption{Dynamic Routing: Even though it is possible to model, the resulting model can be confusing, showing the limits of BPMN.}
\label{bpmn_d_r}
\end{figure}

The size of \autoref{bpmn_d_r}, illustrates how the the dynamic routing pattern can be modeled using BPMN. Again, the message flows can look confusing.

The dynamic routing pattern process consists of four pools: the \emph{Customer, Sales, Warehouse} and \emph{Transport} pool. The process starts with the customer creating and sending a request to the \emph{Sales} pool. This pool is actually only used to forward the request to the warehouse and serves no other purpose. The warehouse receives and processes the request. The dynamic routing takes place then and the \emph{Warehouse} pool either sends a notification to the \emph{Customer} or a shipping order to the \emph{Transport} pool. In the first case, the customer receives the notification and sends a shipping order to the \emph{Transport} pool. The \emph{Transport} pool then sends a shipping notification to the warehouse and the process ends for all participants. In the other case, the warehouse sends a shipping order to the \emph{Transport} pool, which then creates a shipment notification and sends it to the customer, also causing the process to end.

Even though the process, again, may look confusing, it is a valid process model. 

\begin{figure}[htbp]
\centering
\includegraphics[keepaspectratio,width=3.0in,height=0.75\textheight]{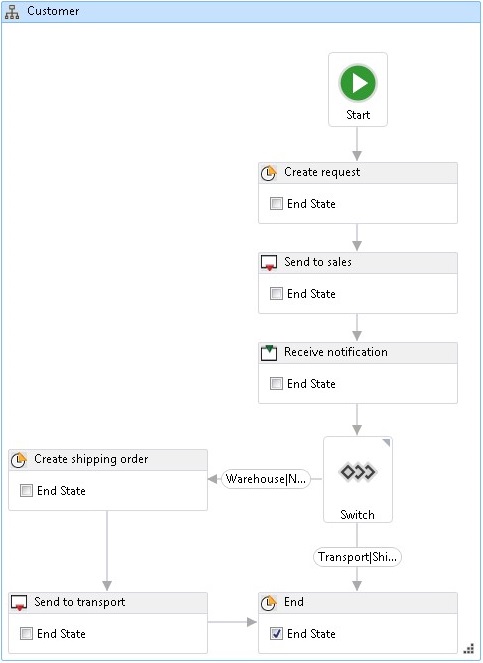}
\caption{Dynamic Routing: The customers actions depend on the received message. The process ends either way, but in one case a message is sent beforehand.}
\label{promi_d_r_c}
\end{figure}

\begin{figure}[htbp]
\centering
\includegraphics[keepaspectratio,width=3.0in,height=0.75\textheight]{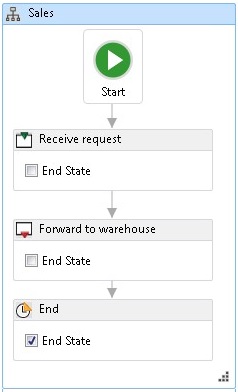}
\caption{Dynamic Routing: The \emph{Sales} subject only has a small role (forwarding a message) and could theoretically be left out of the process.}
\label{promi_d_r_s}
\end{figure}

\begin{figure}[htbp]
\centering
\includegraphics[keepaspectratio,width=3.0in,height=0.75\textheight]{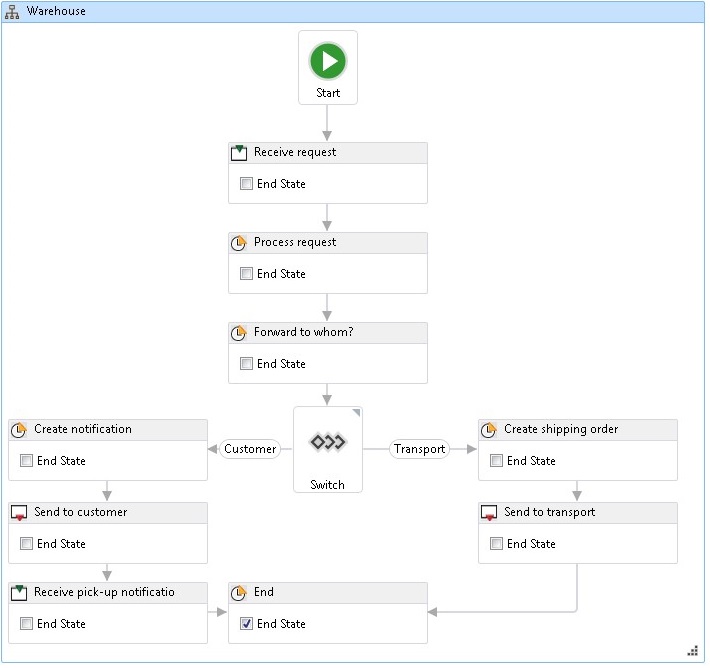}
\caption{Dynamic Routing: The \emph{Warehouse} subject also has two forks: Either the customer or the warehouse is notified, which is effectively dynamic routing. In the first case, with the customer, one last notification has to arrive (sent by the warehouse) before the process can end.}
\label{promi_d_r_w}
\end{figure}

\begin{figure}[htbp]
\centering
\includegraphics[keepaspectratio,width=3.0in,height=0.75\textheight]{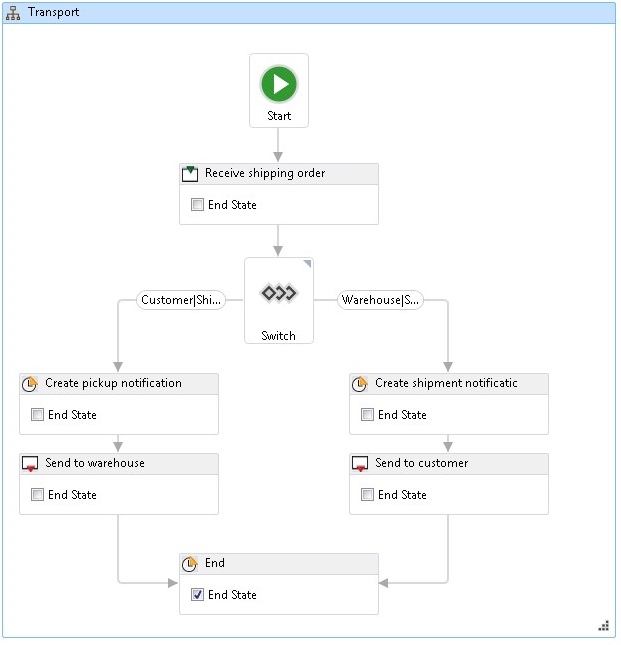}
\caption{Dynamic Routing: The \emph{Transport} subject has two forks and acts on the message which is received. Depending on the inbound message, either the warehouse or the customer receives an outbound message.}
\label{promi_d_r_t}
\end{figure}

With four interacting subjects, this is the most extensive process modeled in \emph{StrICT}. The process itself only uses plain S-BPM. \autoref{promi_d_r_c} shows the \emph{Customer}, \autoref{promi_d_r_s} the \emph{Sales}, \autoref{promi_d_r_t} the \emph{Transport}, and \autoref{promi_d_r_w} the \emph{Warehouse} subject.

The process starts, as is most often the case, with the customer creating a request. This request is then sent to the sales department and the customer waits for a notification.

The sales department only has a small role here, simply receiving the request and forwarding it to the warehouse. 

The warehouse, however, receives the request, processes it and decides what to do (effectively routing dynamically). If the warehouse decides to choose the customer, a notification is created and sent to the customer. The customer receives said notification and creates a shipping order which is sent to the \emph{Transport} subject. The \emph{Transport} subject receives the shipping order, creates a pickup notification and sends it to the warehouse. After that, the process is finished for all participating subjects.

On the other hand, if the warehouse chooses the \emph{Transport} subject instead, a shipping order is created and sent to transport. There a shipment notification is conducted and sent to the customer. After the customer receives it, the process ends for all participating subjects.

\section{Discussion}
\label{discussion}

As mentioned, we were able to model and execute all \emph{Service Interaction Pattern} on our StrICT architecture. We also made an evaluation of some other ``cloud based'' BPMS we could find, based on evaluation versions: \emph{IBM Blueworks Live}, \emph{IYOPRO}, and \emph{ProcessMaker}; we could not include \emph{PegaCloud} and \emph{AppianCloud} in this comparison as they did not answer our requests for an evaluation version. Only \emph{IYOPRO} offered process modeling based on BPMN 2.0. In total \emph{IBM Blueworks Live} could model and execute 6 and \emph{ProcessMaker} 5 \emph{Service Interaction Pattern} (there is a total of 13). \emph{IYOPRO} could model and execute in principle all patterns, but with limitations and workarounds. None of the evaluated platforms offered multi-enterprise functionality to realize cross-enterprise processes.

To sum up: The agent-based and, more specific, the Subject-oriented approach is a natural way to model and execute distributed processes. It is furthermore a promising approach towards a grounded theory of socio-technical systems as it is based on communication as a general concept, not only in computer sciences, but also in social sciences. This has also clear impacts for the development of more natural modeling languages and applications. We could also demonstrate, that the needed technology is available and can be integrated to build a multi-enterprise process platform based on cloud technology.

The platform will be available at the workshop for interactive discussion.

\bibliographystyle{splncs03}
\bibliography{CBI_2014_SIN}

\end{document}